\documentclass[pra,showpacs,amsfonts,amsmath,twocolumn,superscriptaddress,a4paper]{revtex4}
\usepackage{array,amsmath,amsfonts,amssymb,dsfont} \usepackage{natbib}
\usepackage[T1]{fontenc} \usepackage{ae} \usepackage{color}
\usepackage{graphicx}

\begin{document}

\title{Multi-mode mediated exchange coupling in cavity QED}

\author{S.~Filipp} 
\email{filipp@phys.ethz.ch} 
\affiliation{Department
of Physics, ETH Zurich, CH-8093 Zurich, Switzerland}

\author{M.~G\"oppl} 
\affiliation{Department of Physics, ETH Zurich,
CH-8093 Zurich, Switzerland} 

\author{J.~M.~Fink}
\affiliation{Department of Physics, ETH Zurich, CH-8093 Zurich,
Switzerland} 
\author{M.~Baur} 
\affiliation{Department of Physics, ETH
Zurich, CH-8093 Zurich, Switzerland} 

\author{R.~Bianchetti}
\affiliation{Department of Physics, ETH Zurich, CH-8093 Zurich,
Switzerland} 
\author{L.~Steffen} 
\affiliation{Department of Physics, ETH Zurich,
CH-8093 Zurich, Switzerland} 

\author{A.~Wallraff} \affiliation{Department of Physics, ETH Zurich,
CH-8093 Zurich, Switzerland}


\renewcommand{\i}{{\mathrm i}} \def\1{\mathchoice{\rm 1\mskip-4.2mu
l}{\rm 1\mskip-4.2mu l}{\rm 1\mskip-4.6mu l}{\rm 1\mskip-5.2mu l}}
\newcommand{\ket}[1]{|#1\rangle} \newcommand{\bra}[1]{\langle #1|}
\newcommand{\braket}[2]{\langle #1|#2\rangle}
\newcommand{\ketbra}[2]{|#1\rangle\langle#2|}
\newcommand{\opelem}[3]{\langle #1|#2|#3\rangle}
\newcommand{\projection}[1]{|#1\rangle\langle#1|}
\newcommand{\scalar}[1]{\langle #1|#1\rangle}
\newcommand{\op}[1]{\hat{#1}} \newcommand{\vect}[1]{\boldsymbol{#1}}
\newcommand{\id}{\text{id}}

\begin{abstract} Microwave cavities with high quality factors enable coherent coupling of distant quantum systems. Virtual photons lead to a transverse exchange interaction between qubits, when they are non-resonant with the cavity but resonant with each other. We experimentally probe the inverse scaling of the inter-qubit coupling
with the detuning from a cavity mode and its proportionality to the
qubit-cavity interaction strength. We demonstrate that the
enhanced coupling at higher frequencies is mediated by multiple
higher-harmonic cavity modes. Moreover, in the case of resonant qubits,
the symmetry properties of the system lead to an allowed two-photon
transition to the doubly excited qubit state and the formation of a dark
state.
\end{abstract}

\maketitle

\section{Introduction} 
Experiments on single photons coupled strongly to single
 (artificial) atoms \cite{Haroche2007}  allow for in-depth studies of
photon-atom interactions on a single particle level. This has first been demonstrated with individual atoms coupled to microwave 
\cite{Meschede1985,Raimond2001} and later optical cavity fields
\cite{Miller2005,Walther2006}.
In solids, strong
coupling has been achieved with quantum dots
\cite{Reithmaier2004,Yoshie2004} and superconducting circuits
\cite{Wallraff2004b}. Despite the diversity of physical realizations
the coherent exchange of energy between photons and atoms can be
described in all these systems by a generic model named after
Jaynes and Cummings \cite{Jaynes1963}.

In circuit quantum electrodynamics experiments, superconducting
quantum circuits are coupled to single microwave photons in a planar
transmission line cavity \cite{Blais2004}. In this configuration, coupling strengths exceed
decay rates by two orders of magnitude, and strong resonant coupling
between a microwave cavity and a single \cite{Wallraff2004b,
Blais2004,Houck2007,Fink2008,Hofheinz2009} or multiple
\cite{Sillanpaa2007,Fink2009} superconducting qubits has
been observed. In the case of finite detuning between a single
qubit and a resonator mode, energy exchange between the individual
systems is strongly suppressed due to energy conservation. In this
dispersive regime, a residual interaction mediated via virtual photons
 induces a finite Lamb \cite{Fragner2008} and ac-Stark
shift \cite{Schuster2005} of the energy levels. For two qubits coupled
to a common cavity field, the same mechanism leads to an interaction
mediated by virtual photons \cite{Gywat2006} as experimentally
demonstrated \cite{Majer2007}. This coupling is similar to the
J-coupling of interacting nuclear spins as observed in nuclear
magnetic resonance experiments
(e.\,g.\,\cite{Abragam1961,Vandersypen2004}). It is also a dominant interaction of double quantum dots, where the exchange splitting between spin singlet and spin triplet state can be
used to control a logical qubit state encoded in a two-electron
spin-state \cite{Petta2005}. 
 In contrast to these local interactions,
and also opposed to the direct coupling of superconducting quantum
circuits \cite{Yamamoto2003,McDermott2005,Steffen2006, Hime2006,
Niskanen2007, Plantenberg2007, Neeley2010}, the coupling mediated via
virtual resonator photons allows for a long-range interaction between two
or more distant superconducting qubits. In the context of quantum information processing, it can be used to realize two-qubit gates \cite{Blais2007} with superconducting
qubits \cite{DiCarlo2009,DiCarlo2010,Neeley2010}.

In this paper we measure the exchange coupling as a function of
detuning of two qubits from a single or multiple resonator modes and
characterize the symmetry properties of the coupled system. In Section
\ref{sec:Jcoupling} the inter-qubit coupling mechanism and its
spectroscopic measurement is outlined. In Section \ref{sec:singlemode}
the coupling near a single resonator mode is analyzed. In Section \ref{sec:multimode} higher harmonic modes of the transmission line resonator
are included in the analysis. Section
\ref{sec:darkstate} and \ref{sec:twophoton} describe the formation of
a dark state at the avoided level crossing and the observation of a two-photon
transition from the ground to the doubly excited state that is allowed only at qubit resonance.

\section{Exchange coupling mechanism}
\label{sec:Jcoupling}

\begin{figure} \centering
  \includegraphics[width=86mm]{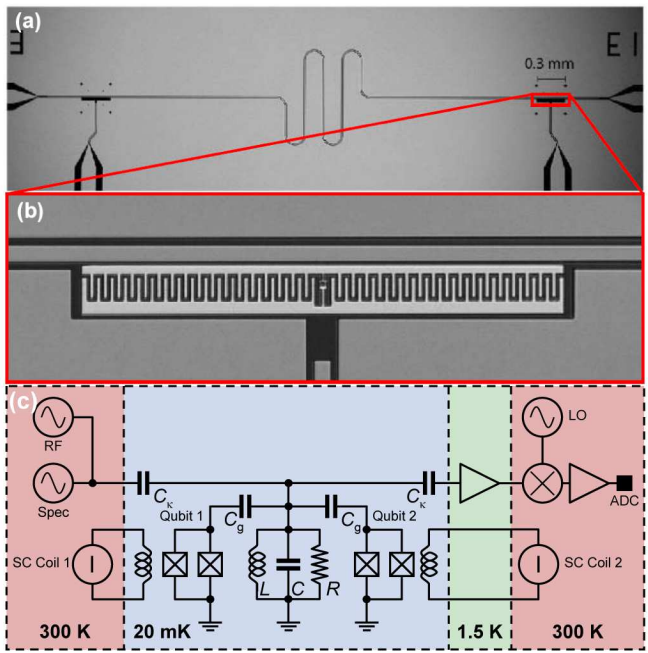}
  \caption{(a) Coplanar waveguide resonator coupled via finger
capacitors $C_\kappa$ to input and output transmission lines. Two
transmons are capacitively coupled to the resonator at its ends
($C_g$). Additional ac-signal lines are capacitively coupled to the
qubits (not used in the experiments). (b) Optical micrograph of a
transmon qubit.  (c) Schematics of the measurement setup.  The state
of the qubit is determined by measuring the transmission of the RF
signal through the transmission line cavity modeled as an LCR
oscillator.  When a spectroscopy signal (Spec) is resonant with a
qubit transition, the resonance frequency of the cavity is shifted and
the change in transmission amplitude is recorded at the analog-digital converter (ADC) after
down-conversion with a local oscillator (LO) \cite{Bianchetti2010}. The
qubit frequencies can be tuned independently with superconducting
coils (SC Coil 1/2).}
\label{fig:setup}
\end{figure}

In our experiments two superconducting qubits are dispersively coupled
to a microwave cavity, see Fig.~\ref{fig:setup}. The quantum
circuits are realized as weakly anharmonic transmon qubits
\cite{Koch2007} and the cavity is formed by a $\lambda/2$
coplanar-waveguide resonator supporting several harmonic modes
\cite{Goppl2008}.
In the dispersive regime, the detuning $\Delta_j^{(i)} \equiv
\omega_{ge}^{(i)} - \omega_j$ is larger than the coupling strength
$g_j^{(i)}$ of both qubits ($i=1,2$) to each resonator mode $j$. 
%
The relevant Hamiltonian
\begin{align}
\label{eq:HJ}
 H_J = &\hbar\sum_{i=1,2} \frac{\omega_{ge}^{(i)}}{2}
\sigma_z^{(i)} + \hbar\sum_j (\omega_j + \chi_j^{(1)} + \chi_j^{(2)}) a_j^\dagger a_j + \\
\nonumber
&+\hbar J \left(\sigma^{(1)}_+\sigma^{(2)}_- +
\sigma^{(2)}_+\sigma^{(1)}_-\right)
\end{align} 
is obtained by adiabatically eliminating the direct
 qubit-resonator interaction of the qubits 
for each harmonic mode $a_j$ in the Jaynes-Cummings Hamiltonian \cite{Blais2007}.  The first term denotes
the qubit Hamiltonian with Lamb-shifted transition frequencies
$\omega_{ge}^{(i)}$  from the ground to the first excited
state. Higher transmon levels do not play a role in our experiments
and are therefore neglected. 
The second term in Eq.~(\ref{eq:HJ})
describes the resonator modes with frequencies $\omega_j = (j+1)
\omega_0$, integer multiples of the fundamental frequency
$\omega_0$, shifted by the cavity pulls $\chi_j^{(i)}$ \cite{Blais2007,Filipp2009b}. Finally, the third term describes the effective
qubit-qubit coupling, also called J-coupling or transverse exchange
coupling, 
\begin{equation}
\label{eq:J} J = \frac{1}{2} \sum_j
g_j^{(1)}g_j^{(2)}\left(\frac{1}{\Delta_{j}^{(1)}} +
\frac{1}{\Delta_{j}^{(2)}}\right),
\end{equation} a flip-flop interaction mediated by virtual
photon exchange.  

\begin{figure*}[htpb] \centering
  \includegraphics[width=180mm]{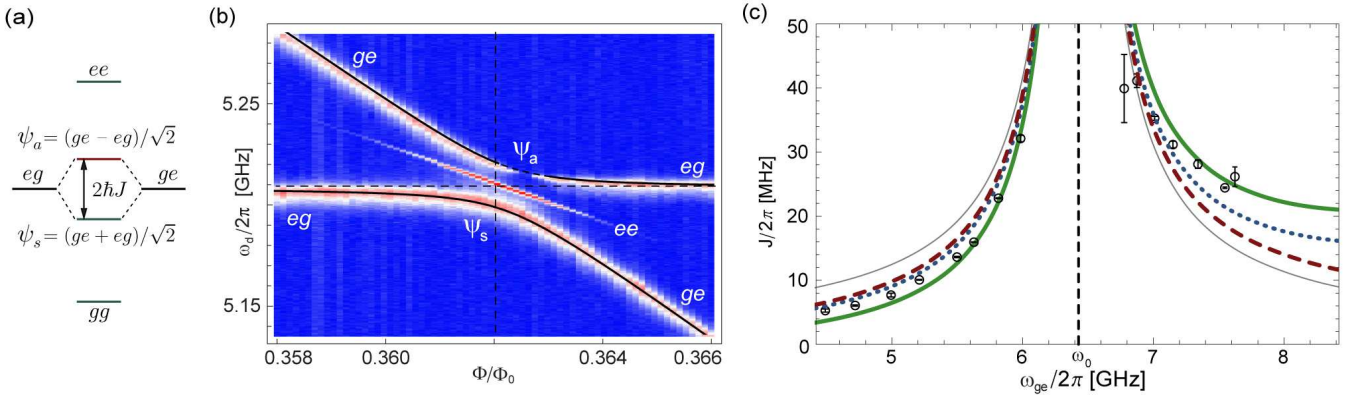}\\ 
\caption{(a) Energy level diagram of two transversely coupled transmon qubits. (b)
Spectroscopic measurement of the avoided level crossing in sample A as function of
normalized flux $\Phi/\Phi_0$ threading the first qubit loop with the
second qubit at a fixed frequency. The solid lines indicate energy
levels calculated from a diagonalization the two-qubit Jaynes-Cummings
Hamiltonian. (c) Experimentally extracted value of the coupling
strength J as a function of qubit frequency (dots). Lines indicate
calculated values of $J(\omega_{ge})$ for different models, see text for
details.}
\label{fig:2Qsplitting}
\end{figure*}

The transverse exchange coupling  in Eq.~(\ref{eq:J}) leads to an
avoided level crossing of the excited qubit states \cite{Majer2007}.
At qubit-resonance, where $\delta_q \equiv
\omega_{ge}^{(1)}-\omega_{ge}^{(2)}=0$, the size of the splitting is
$2J = \hbar\sum_j 2 g_j^{(1)}g_j^{(2)}/\Delta_j$. The new eigenstates
are the symmetric triplet states $\ket{gg}$, $\ket{ee}$ and
$\ket{\psi_s} = (\ket{ge} + \ket{eg})/\sqrt{2}$, as well as the
anti-symmetric singlet state $\ket{\psi_a} = (\ket{ge} -
\ket{eg})/\sqrt{2}$, see Fig.~\ref{fig:2Qsplitting}(a). In the
maximally entangled states $\ket{\psi_{s/a}}$ a single excitation is
shared between the two qubits. More generally, for $\delta_q\neq 0$ the eigenstates of
the Hamiltonian in Eq.~(\ref{eq:HJ}) can be parametrized as
\begin{align}
\label{eq:psipm} \ket{\psi_s}&= \sin\theta_n \ket{ge} + \cos\theta_n
\ket{eg},\\ \nonumber \ket{\psi_a}&=\cos\theta_n \ket{ge} -
\sin\theta_n \ket{eg},
\end{align} with the mixing angle $\theta_n$ determined by $\cos
2\theta_n = -\delta_q/\sqrt{4J^2+\delta_q^2}$ and $\sin
2\theta_n = 2J/\sqrt{4 J^2 + \delta_q^2}$.  The separable
qubit states $\ket{eg}$ and $\ket{ge}$ are asymptotically realized,
$\ket{\psi_a} \rightarrow\ket{eg}$ and
$\ket{\psi_s}\rightarrow\ket{ge}$, for large qubit-qubit detunings
($\delta_q \rightarrow \infty$), as indicated in
Fig.~\ref{fig:2Qsplitting}(b).


We have performed two sets of experiments using samples with different parameters listed in table \ref{tab:parameters}.
In these experiments, the energy spectrum of the coupled qubits is probed
by monitoring the transmission through the resonator while applying a
second spectroscopy tone \cite{Schuster2005} at frequency $\omega_d$. For the spectroscopy
measurement shown in Fig.~\ref{fig:2Qsplitting}(b), the first qubit is
kept at a fixed frequency $\omega_{ge}^{(1)}/2\pi$ and the second qubit
frequency $\omega_{ge}^{(2)}/2\pi$ is swept across the avoided
crossing by changing its flux bias using external
coils. The value of $J$ can be extracted from a fit of the upper and
lower branch of the avoided crossing to the function
\begin{equation}
f(\omega;\omega_{ge}^{(1)},J) = \left((\omega+\omega_{ge}^{(1)})\pm
\sqrt{(\omega_{ge}^{(1)}-\omega)^2+4J^2}\right)/2,
\end{equation} 
where $\omega$ is in this parameter regime an approximately linear function 
of the flux $\Phi$ threading the
second qubit loop.  The fit parameters are the transition frequency
$\omega_{ge}^{(1)}$ of the first qubit and the coupling strength
$J$. Both are determined with a precision of typically better than
$0.5~\rm{MHz}$. In this particular example we find
$\omega_{ge}^{(1)}/2\pi = 5.210\pm0.00005~\rm{GHz}$ and $J/2\pi =
10.06\pm0.06~\rm{MHz}$.

Two additional features are observed in
Fig.~\ref{fig:2Qsplitting}(b). First, a third spectroscopic line
centered between the upper and the lower branch appears at higher
drive powers. This is a signature of a two-photon 
transition from the ground state ($\ket{gg}$) to the doubly excited state ($\ket{ee}$) of the coupled qubit system that is only allowed directly
at the anti-crossing. This is discussed in Section
\ref{sec:twophoton}. Second, the upper branch shows a transition to a
dark resonance at the avoided crossing, which can be explained by the
symmetry of the states with respect to the spectroscopic drive, see
Section \ref{sec:darkstate}.

\section{Coupling to the fundamental resonator mode}
\label{sec:singlemode}

According to Eq.~(\ref{eq:J}) the coupling $J = \hbar
g^{(1)}g^{(2)}/\Delta$ scales inversely with the detuning $\Delta =
\Delta^{(1)}=\Delta^{(2)}$, considering only a single resonator
mode. We have recorded the avoided crossing between the two qubits at
different detunings $\Delta$ from the fundamental mode of the
resonator using a spectroscopic measurement performed on sample A. The
corresponding parameters are listed in Table \ref{tab:parameters}. The
measured values of $J$ shown in Fig.~\ref{fig:2Qsplitting}(c) are
determined for each detuning from a fit as described in Section
\ref{sec:Jcoupling}.
\begin{table}[b] \centering
  \begin{tabular}{p{27mm}p{27mm}p{23mm}} Parameter & Sample A & Sample
B\\\hline\hline $\omega_0/2\pi$ & 6.44 GHz & 3.34 GHz\\
$\kappa/2\pi$ &  1.57 MHz & 1.91 MHz\\
$E_C^{(1)}/h$ & 232 MHz & 148 MHz\\ $E_C^{(2)}/h$ & 233 MHz &
153 MHz\\ $E_J^{(1)}/h$ & 35 GHz & 409 GHz\\ $E_J^{(2)}/h$ & 38
GHz & 375 GHz\\ $g^{(1)}_0/2\pi$ & 133 MHz & 43 MHz\\ $g^{(2)}_0/2\pi$ &
134 MHz & 42 MHz\\\hline
  \end{tabular}
  \caption{Parameters of samples A and B as determined from
independent measurements. $\omega_{0}$ denotes the fundamental frequency, $\kappa$ the cavity decay rate, $E_C^{(1,2)}$ the charging energy, $E_{J}^{(1,2)}$ the
maximum Josephson energy and $g^{(1,2)}_0$ the coupling strength to the
fundamental cavity mode of qubits $1$ and $2$. }
  \label{tab:parameters}
\end{table}

Considering only one relevant resonator mode and constant $g$, the
strength of the inter-qubit coupling is expected to be symmetric about
the resonator frequency (Eq.~(\ref{eq:J}); thin gray line in
Fig.~\ref{fig:2Qsplitting}(c)). The asymmetry in the data can partly
be accounted for by including the frequency-dependence of the coupling $g$, as explained in
Appendix \ref{sec:gvsflux}. It follows from the transition matrix elements that $J$ scales proportional to the
transition frequency $\omega_{ge}$ of the qubits. This scaling factor leads to  an asymmetry of the exchange coupling around the resonance
frequency $\omega_0$ 
that improves the agreement with the data (dashed red line, Fig~\ref{fig:2Qsplitting}(c)).
To check, whether the remaining discrepancy originates from the
dispersive approximation, we have also done a numerical
diagonalization of the full generalized Jaynes-Cummings Hamiltonian
(not shown). 
This calculation agrees with the dispersive model within the errors of the experimentally determined values of $J$.

\begin{figure} 
\centering
  \includegraphics[width=86mm]{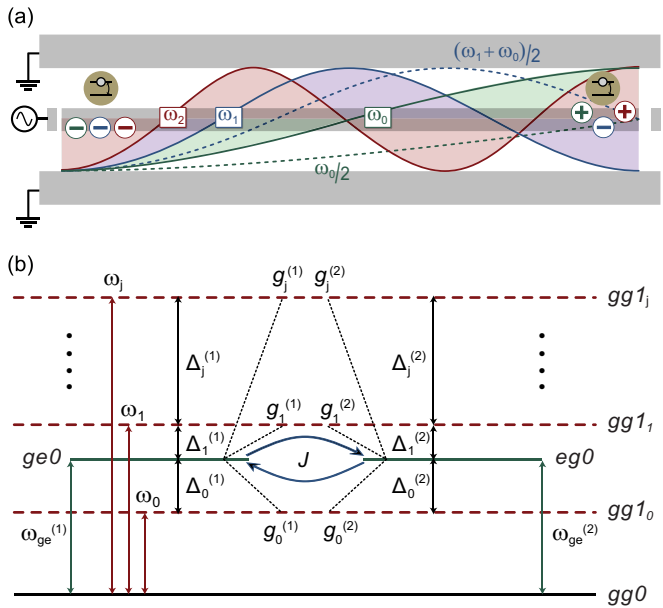}
  \caption{(a) Spatial mode structure of a coplanar waveguide
resonator. Qubits are positioned at opposite ends of the $\lambda/2$
resonator with coupling of alternating sign, $g_j^{(1)} =
(-1)^{j+1}g_j^{(2)}$ to the $j$-th resonator mode at frequency
$\omega_j$. (b) Energy level diagram and coupling scheme for two
qubits with transition frequencies $\omega_{ge}^{(1)}$ and
$\omega_{ge}^{(2)}$ coupled to a transmission line cavity with
fundamental frequency $\omega_0$. Energy levels with a photon in the
$j$-th resonator mode ($\rm{gg}1_{\rm{j}}$) or a qubit excitation
($\rm{ge0}$ or ${eg0}$) are shown. The exchange interaction $J$
depends on the detunings $\Delta_j^{(i)}$ and the coupling strengths
$g_j^{(i)}$ of both qubits $i=1,2$ to the resonator mode $j$.}
  \label{fig:couplingscheme}
\end{figure}

For a quantitative agreement, higher harmonics of the resonator
have to be considered.  The particular implementation of the resonator
as an open-ended coplanar waveguide supports higher harmonics at
integer multiples of the fundamental frequency \cite{Goppl2008}, see
Fig.~\ref{fig:couplingscheme}(a).  Each of these higher modes provides
a channel for the exchange of virtual photons between the qubits
determined by the detuning $\Delta^{(i)}_j$ and the coupling
$g^{(i)}_j$ to the harmonic mode $j$ as indicated in
Fig.~\ref{fig:couplingscheme}(b).  Above the fundamental mode, the
coupling to the first harmonic mode $j=1$ contributes significantly to the qubit-qubit coupling, which results in an asymmetry with respect to the detuning $\Delta$. Including four
modes in Eq.~(\ref{eq:J}) to determine the expected value of $J$, good
agreement with data is obtained (thick green line,
Fig.~\ref{fig:2Qsplitting}(c)). 

It is important to also include the
alternating sign of the electric fields at the qubits' position in the
calculations. As depicted schematically in
Fig.~\ref{fig:couplingscheme}(a), the electric field of the
fundamental and higher even modes (j=0,\,2,\,4,\ldots) have always
opposite sign, i.~e. a relative phase of $\pi$, at either end of the
microwave cavity, whereas odd modes (j=1,\,3,\,5,\ldots) have equal
sign. Thus, higher harmonics add with different signs to the effective
coupling strength.  A priori, the sum in Eq.~(\ref{eq:J}) has to be
extended over all modes.  The sum does, however, not
converge, as discussed also in the context of
Purcell-limited qubit decay rates in \cite{Houck2008}.  The proportionality of the coupling $g^{(1)}_j g^{(2)}_j
\propto \omega_j = (j+1) \omega_0$ (Eq.~\ref{eq:gscale}) together with
the same proportionality of the detuning $\Delta_j\propto
(j+1)\omega_0$ for large $j$ leads to an non-converging series
alternating between the two values
$$J_{\rm{even}}= \hbar \sum_{j=0}^{2k} (-1)^{j+1}
\frac{g_j^2}{\Delta_j}\ \ \mbox{and}\ \ J_{\rm{odd}} 
= \hbar \sum_{j=0}^{2k+1} (-1)^{j+1}
\frac{g_j^2}{\Delta_j}.$$
Apparently, a cut-off frequency has to be imposed to obtain physical
results.
In Fig.~\ref{fig:2Qsplitting}(c) we have also included a plot of $J_{\rm{even}}$
when terminating the sum at the fourth harmonic ($k=2$) (dotted blue
line). It is observed that the difference between an even and odd
number of modes is significant, up to $25\%$ of the coupling
strength. We have also verified that this is not an artifact of the
dispersive model by numerically diagonalizing the
Jaynes-Cummings Hamiltonian.

In our measurements we observe enhancement of the exchange coupling, inversely proportional to the detuning of the qubits to the cavity mode. The asymmetry around the mode is attributed to higher harmonic modes that contribute to the measured (renormalized) $J$. However, to compute the coupling strength the number of included modes has to be restricted by imposing a high-frequency cut-off. Physically, there are several  mechanisms conceivable. The energy needed to overcome the pairing
interaction of Cooper pairs  sets an upper frequency of about
$700~\rm{GHz}$ for Niobium. 
 The electric field across the transmon averages out
when the wavelength of the photons becomes comparable to the size of
the transmon at a frequency of about $400~\rm{GHz}$.  
Also, radiation or dielectric loss mechanisms and the photon loss rate
through the coupling capacitors increase at higher
frequencies \cite{Houck2008}. Current experiments are, however, not designed to work at
frequencies higher than approx.~$15~\rm{GHz}$. The exploration of the relevant frequency range will require an elaborate circuit architecture and will be challenging with current technology.

\section{Multi mode coupling}
\label{sec:multimode}

\begin{figure*}[htpb] \centering
    \includegraphics[width=172mm]{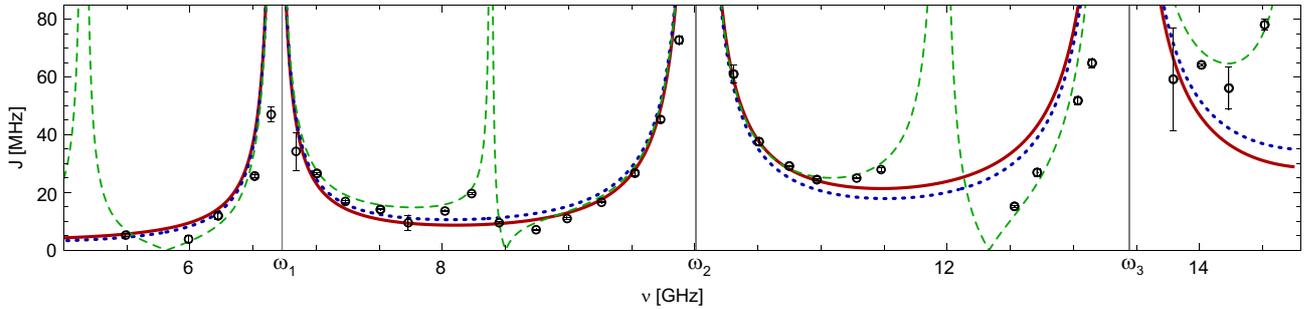} 
\caption{Exchange coupling $J$ versus qubit frequency in sample B. The vertical lines indicate the frequency
of the coplanar waveguide modes $\omega_j$. Experimental data (dots) confirm the expected increase of
$J$ with increasing frequency. The solid red  (dotted blue) line indicates the
calculated $J$ including $N=6$ ($N=7$) resonator modes. The dashed
green line is a fit to a model
with additional resonances.} \label{fig:multimode}
\end{figure*}
To assess the coupling to higher-order modes, we measure the qubit-qubit coupling strength $J$ as a function of qubit
transition frequency  in a second
sample B over a broader frequency range. In this sample
the frequency of the fundamental resonator mode is lower,
$\omega_0/2\pi = 3.34~\rm{GHz}$, and the maximal Josephson energy
$E_{J}$ is higher by about one order of magnitude (see Table
\ref{tab:parameters} - sample B). As a result, the qubit transition
frequency can be swept over several resonator modes. 

The measured
inter-qubit coupling strength shows an enhanced value around each harmonic
mode, as well as an overall increase  with
frequency (Fig.~\ref{fig:multimode}) in agreement with the discussion
in Sec.~\ref{sec:singlemode}.  A calculation based on the dispersive Jaynes-Cummings
Hamiltonian can explain the data
qualitatively.  However, including an even ($N=6$, solid red line) or an
odd ($N=7$, dotted blue line) number of resonator modes yields again
significant differences in the calculated value of $J$, neither of
the curves resulting in good quantitative agreement.
%
%
%

The resonator modes discussed so far cannot fully explain the measured
inter-qubit coupling. Large deviations located asymmetrically around $\approx 8.5~\rm{GHz}$ and
$\approx 12~\rm{GHz}$ (Fig.~\ref{fig:multimode}) hint at the presence of an
additional coupling mechanism at these intermediate frequencies. The
measurement reveals that the coupling strength is asymmetrically
modified between every two resonances alluding to an anti-resonance
that mediates a qubit-qubit coupling channel of similar magnitude as
the coplanar waveguide resonance. This can lead to an enhancement or
suppression of the qubit-qubit coupling at intermediate frequencies
due to interference between multiple resonances.  

To account for these spurious resonances, additional field modes are
included in the model. Physically, these modes may be identified as
the slotline modes of the coplanar waveguide, a differential
excitation of the left and right ground plane
\cite{Wolff2006}. Air-bridges or wire-bonds that connect the
ground planes could effectively suppress these modes,
but have not been implemented in this sample.
Treating these modes equivalently to the coplanar
waveguide modes in the derivation of the dispersive multi-mode
Jaynes-Cummings Hamiltonian (\ref{eq:HJ}), an extra contribution to the exchange coupling
emerges,
$$J_{\rm{tot}} = J + \tilde{J} = J + \frac{1}{2} \sum_l^M \tilde{g}^{(1)}_l \tilde{g}^{(2)}_l \left(\frac{1}{\tilde{\Delta}_{l}^{(1)}} +
  \frac{1}{\tilde{\Delta}_{l}^{(2)}}\right).$$

$J_{\rm{tot}}$ is then fitted to the data in Fig.~\ref{fig:multimode}
assuming that the coupling strengths of both qubits are equal in
magnitude, $|\tilde{g}^{(1)}_l| = |\tilde{g}^{(2)}_l|$, but have --
like the coplanar waveguide modes -- alternating sign,
$|\tilde{g}^{(1)}_l| = (-1)^{l+1}|\tilde{g}^{(2)}_l|$ with
$l=0,1,\ldots$. For the fit we take four extra modes into account
($l=1,2,3,4$) from which we obtain the resonance frequencies 
$\tilde{\omega}_{1,2,3,4}/2\pi \approx
\{5.2,\,8.4,\,11.9,\,14.8\}~\rm{GHz}$. 
Note, that these additional modes are
not observed in simple transmission measurements \cite{Goppl2008} of
the resonator and their frequencies can therefore not be determined independently.   In the same fit the ratios
$\tilde{g}_j/g_j$  are determined to
$\tilde{g}_{1,2,3,4}/g_{1,2,3,4}=\{1(1),\,0.5(2),\,0.8(2),\, 0.7(1)\}$, which
shows that the coupling strengths to the spurious modes $\tilde{g}_i$ and to the coplanar waveguide modes $g_i$ are similar in strength. This similarity hints at a highly localized field of the spurious mode with
 small effective mode volume. Also, the relative
sign between the couplings $\tilde{g}_l^{(1)}$ and $\tilde{g}_l^{(2)}$ of the qubits to 
the spurious mode alternate with the mode number $l$. This implies that the field of the spurious modes has also
either equal or opposite direction at the position of the qubits, like
the coplanar waveguide mode outlined in Fig.~\ref{fig:couplingscheme}. The qualitative agreement to the measured values of J is
considerably improved by the inclusion of these extra modes, see dashed green line in
Fig.~\ref{fig:multimode}.  

The measured values of the exchange coupling $J$ demonstrate the sensitivity of the
qubit-qubit coupling to the full mode structure of the circuit.   While a
single-mode model is sufficient around a single resonance,
quantitative predictions require complete knowledge of designed and
spurious resonances. Appropriate circuit design and use of wire-bond
or air-bridge
connections of ground-planes on the chip can
short out spurious modes. In contrast, additional resonances can also
be incorporated on purpose into the circuit design \cite{Reed2010} to
modify the qubit-qubit coupling at certain frequencies.

 \section{Dark state}
\label{sec:darkstate} 

A characteristic feature of the avoided level
crossing is the observation of a dark resonance, where
the transition from the ground state to the upper energy branch is forbidden and no signal
is observed in  spectroscopy measurements, see Fig.~\ref{fig:2Qsplitting}(b). In fact, the
symmetry of the states at the avoided crossing leads to a selection
rule with respect to the spectroscopic drive
\cite{Blais2007,Majer2007}
\begin{equation}
\label{eq:drive} H_d = \epsilon
\left(g^{(1)}\sigma_+^{(1)}/\Delta^{(1)} +
g^{(2)}\sigma_+^{(2)}/\Delta^{(2)}\right) + h.c.
\end{equation} through the resonator. To see this, we decompose the
eigenvalues of the dispersive Hamiltonian (\ref{eq:HJ}) into 
triplet states and a singlet state. These are the eigenstates of the
permutation operator (Eq.~\ref{eq:permop})
to the 
eigenvalue $\pm 1$. Explicitly, the triplet states $\ket{gg}$ (ground state),
$\ket{ee}$ (doubly excited state) and the symmetric state $(\ket{ge} +
\ket{eg})/\sqrt{2}$  span the symmetric subspace,
whereas the singlet state $(\ket{ge} - \ket{eg})/\sqrt{2}$ is
anti-symmetric under permutation of the qubits.  This is equivalent to
a decomposition of the system into a spin-1 and a spin-0 particle.
The sign of the coupling constants $g^{(i)}$ determines the drive
symmetry. Equal (opposite) sign of the electric field at the position
of the two qubits leads to the positive (negative) sign of the second
term in Eq.~(\ref{eq:drive}) and, consequently, to a(n) (anti-)symmetric
excitation.  For equal sign of the couplings, the drive term and the
permutation operator commute (symmetric drive). Then, only transitions
between states of same symmetry are allowed (see Appendix
\ref{sec:darksymm}) and the anti-symmetric state stays dark at zero
detuning. Vice versa, for opposite sign of the couplings the now
anti-symmetric drive can connect symmetric to anti-symmetric states.

In our experiments the symmetry of the drive is determined by the
frequency of the microwave signal and the distance $d$
between the qubits. The relative sign of the field at the qubit
positions $x_i$ is determined by the phase difference
$\Delta\phi=\phi(x_1) - \phi(x_2) = \omega_d d/c_{\rm{eff}}$ of the
travelling wave between the qubits. Here, we have used the dispersion
relation $k_d = \omega_d/c_{\rm{eff}}$ with the propagation velocity
$c_{\rm{eff}}$ of light in the transmission line. As the qubits are
located at the end of the transmission line resonator, $d$ is
approximately the length of the resonator and sign changes happen at
frequencies $\omega_s= s \pi c_{\rm{eff}}/(2d)$ with $s=1,3,5,\ldots$,
in between two resonances as indicated by the dotted lines in
Fig.~\ref{fig:couplingscheme}(a).

\begin{figure} 
   \centering
   \includegraphics[width=86mm]{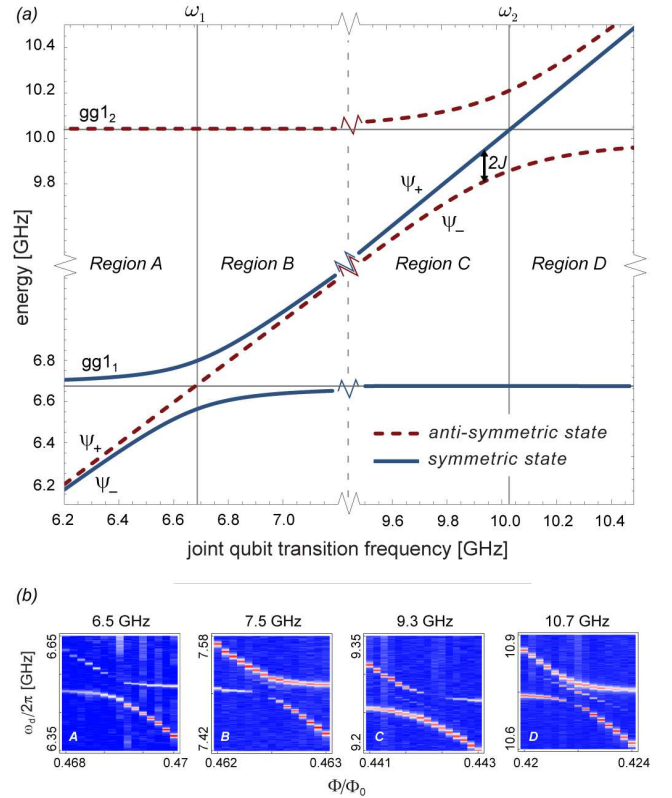}
   \caption{(a) Energy levels of the qubits-resonator system when the
qubits are degenerate and their transition frequencies are
simultaneously swept across the resonator modes. The parameters are
those of sample B (Table \ref{tab:parameters}). (b) Spectroscopic
measurements of the anti-crossing in the  regions labeled A,\,B,\,C
and D in
(a).}
   \label{fig:darkstates}
 \end{figure}

Whether the eigenstate with lower or higher energy is dark, depends --
in the simplest model with only a single dominant resonator mode -- on
the qubit-resonator detuning. The higher ($\ket{\psi_+}$) and lower
($\ket{\psi_-}$) energy eigenstate of the Hamiltonian (\ref{eq:HJ}) at the avoided crossing can according to
Eq.~(\ref{eq:psipm}) be written as $\psi_{\pm} = \left(\ket{ge} \pm
{\rm sign}(J) \ket{eg}\right)/\sqrt{2}$. Note, that in this notation the
subscript denotes higher ($+$) or lower ($-$) energy and not the
symmetry of the state. In Fig.~\ref{fig:darkstates} the energy
levels of the coupled resonator-qubits system are plotted as a
function of the qubit transition frequencies, which are kept
equal, along with their respective symmetries. For a coupling to the \emph{first harmonic} mode $J =
g^{(1)}_1 g^{(2)}_1/\Delta_1 < 0$ below the mode ($\Delta_1<0$,
$g^{(1)}_1 g^{(2)}_1>0$) and $J>0$ above the mode ($\Delta_1>0$,
$g^{(1)}_1 g^{(2)}_1>0$). Consequently, \emph{below} the first
harmonic the higher energy state $\psi_+$ is the anti-symmetric
singlet state. This state cannot be excited from the ground state with
a symmetric drive and stays dark (Fig.~\ref{fig:darkstates}(b) -
sub-panel A). At the avoided crossing with a resonator mode the
lower and higher energy qubit state swap their symmetry due to the
sign change of the detuning. Since the drive does not change its
symmetry, also the dark and bright state energies are interchanged and
the dark state appears at the lower branch (Fig.~\ref{fig:darkstates}(b) -
sub-panel B). The dark state is always closer in
frequency to the resonator transition.

If the \emph{second harmonic} mode dominates the coupling, the
situation is reversed since the coupling constants have different
signs. Below this mode, the $\psi_+$ (higher energy) state is
symmetric, and $\psi_-$ (lower energy) is anti-symmetric. Still, the
dark state appears at the upper branch of the avoided crossing (Fig.~\ref{fig:darkstates}(b) -
sub-panel C) and switches to the
lower branch above the resonator (Fig.~\ref{fig:darkstates}(b) -
sub-panel D). The reason is that the
drive symmetry changes as well in between two resonator modes as
explained above, implying that the drive field changes from a
symmetric (around the first harmonic mode) to an anti-symmetric drive (around
the second harmonic mode). The drive can then induce transitions between
the ground and an anti-symmetric state, but not to the symmetric state
$\psi_+$ of the upper branch.  The various conditions leading to the
identification of the dark state branch are summarized in table
\ref{tab:symmetries}.  In particular, if the
drive (line 4) has different symmetry than the higher energy state
$\psi_+$ (line 3), $\psi_+$ remains dark.

\begin{table}[h] \centering \tabcolsep3mm
  \begin{tabular}{lcccc} Region: & A & B & C &
D\\[1mm]\hline\hline\\[-2mm] $g^{(1)}_j g^{(2)}_j$ & $\oplus$ &
$\oplus$ & $\ominus$ & $\ominus$\\[2mm] $\Delta^{(1)}_j =
\Delta^{(2)}_j$ & $\ominus$ & $\oplus$ & $\ominus$ & $\oplus$\\[2mm]
$J = \Delta^{(1)}_j \Delta^{(2)}_j/\Delta_j$ & $\ominus$ & $\oplus$ &
$\oplus$ & $\ominus$\\ $\ \ \hat{=}\ \psi_+$ symm.&&&&\\[2mm] drive
symm. & $\oplus$ & $\oplus$ & $\ominus$ & $\ominus$\\[2mm] dark state
& $\psi_+$ & $\psi_-$ & $\psi_+$ & $\psi_-$\\ \hline
  \end{tabular}
  \caption{Symmetry considerations leading to the dark state at the
lower or upper energy branch of the avoided level crossing. $\oplus$
($\ominus$) denotes a positive (negative) value or symmetry.}
  \label{tab:symmetries}
\end{table}
%

\section{Two-photon transition}
\label{sec:twophoton}

The spectroscopic line between upper and lower branch of the
avoided level-crossing in Fig.~\ref{fig:2Qsplitting}(b) is a two-photon transition from the ground state $\ket{gg}$ to the doubly excited
state $\ket{ee}$. Similarly, this transition has also been observed in a phase qubit coupled coherently to a two level fluctuator in the tunnel barrier of the Josephson junction comprising the qubit \cite{Bushev2010} and in
molecular spectroscopy of two nearby molecules \cite{Hettich2002}. It
becomes visible only at the center of the avoided level crossing,
again a manifestation of the symmetry properties of the system. 
 The rate of the corresponding two-photon transition \cite{Loudon2000}
$$\Gamma = \frac{2\pi}{\hbar^4}\left|\sum_m \frac{\bra{ee} H_d \ket{m}\bra{m} H_d \ket{gg}}{\omega_{m}-\omega_d}\right|^2\delta(\omega_{ee} - 2\omega_d)$$
is given by a sum over the intermediate states $m$.
Off
the avoided crossing the qubits are effectively decoupled. In this case the intermediate states are $m=ge,eg$. The
transition is then prohibited due to destructive interference between
the two possible paths, $gg\leftrightarrow eg \leftrightarrow ee$ or
$gg\leftrightarrow ge \leftrightarrow ee$, connecting the ground to
the doubly excited state.
Due to the opposite sign of the detunings $\omega_{ge/eg} - \omega_d$, the two terms in the sum cancel and the transition rate $\Gamma = 0$.
With the qubits at resonance, the intermediate states are
$m=\psi^+,\psi^-$ and one term in the sum vanishes due to the
forbidden transition to the dark state. With only one possible path
connecting the $gg$ to the $ee$ state, no interference takes place and
the transition becomes allowed. The enhanced transition rate can be
employed for directly creating the maximally entangled state
$\ket{\phi} = (\ket{gg} + \ket{ee})/\sqrt{2}$.

\section{Conclusion} 
We have analyzed the coupling between two distant
qubits mediated by the harmonic modes of a resonator. We have observed an
overall increase of the exchange coupling with frequency as expected
from a model including higher-harmonic modes of the coplanar waveguide
resonator. Good qualitative agreement over a wide frequency range between the dispersive model and
experimental data is obtained when taking spurious resonances of the
coplanar waveguide in addition to the coplanar waveguide modes into
account. Hence, measurements of the transverse inter-qubit coupling can be employed to
detect and investigate spurious global coupling channels between
distant qubits, complementary to measurements of single qubit spectra
used to detect spurious local resonances
\cite{Simmonds2004,Martinis2005}. How many higher harmonic modes to include in the
theory, i.~e. where to set a high-frequency cut-off, can, however, not be decided on the basis of current measurements. 

In addition, we have observed dark states and enhanced
two-photon absorption at the avoided level crossing in spectroscopic
measurements. These characteristic features are based on the
relation between the symmetry of the drive and the singlet and triplet states formed by the coupled qubits, which also explains the dark state at either the lower or higher energy
branch.
These symmetries also affect decay processes of singlet and triplet
states and, together with the non-trivial environment formed by the
microwave resonator, dissipative dynamics of separable and entangled
states can be studied.  The exchange coupling  can also be useful
for building two-qubit gates when fast flux-pulses are applied to tune
the qubits into resonance. In the context of quantum information
processing, the resulting SWAP gate forms a universal two-qubit gate
with short operation times. Moreover, the transverse exchange coupling
mechanism described in this article mediates interaction not only
between two, but an arbitrary number of distant qubits, an interesting
playground for studies of collective phenomena with superconducting
circuits, where the interaction is not restricted to
nearest-neighbours.

The authors acknowledge useful discussions with A. Blais. Also, 
the group of M. Siegel is acknowledged for the preparation of Niobium films.  This work was supported by
Swiss National Science Foundation (SNF), Austrian Science Foundation
(FWF) and ETH Zurich. 

\appendix
\section{Frequency-dependence of the qubit-resonator coupling}
\label{sec:gvsflux}

The coupling $g$ to the resonator is proportional to the rms voltage
fluctuations of the vacuum field $V_{\rm{rms}}^0$ at the position of
the $i$-th qubit and to the off-diagonal matrix element
\cite{Blais2004,Koch2007} of the charge operator $\op{n}$,
\begin{equation}
\label{eq:g} g = 2 e\beta V_{\rm{rms}}^0 \langle g | \op{n} | e
\rangle.
\end{equation} 
The prefactor $\beta$ is determined by the geometry of
the circuit used in our experiments.  In the large $E_J/E_C$ limit,
realized in the devices, the matrix element is proportional to the
square-root of the qubit transition frequency, $\langle g | \op{n}|
e\rangle \propto \sqrt{\omega_{ge}}$. The vacuum field
$V_{\rm{rms}}^0$ is proportional to the square root of the mode
frequency, $\sqrt{\omega_j} = \sqrt{(j+1) \omega_0}$
\cite{Blais2004,Haroche2007}. 

For a single qubit on
resonance with the $j$-th resonator mode, $\omega_{ge} = (j+1)
\omega_0$, the scaling of the qubit-resonator coupling is approximately linear in
the mode number, $g \propto (j+1)\omega_0$. To verify the linearity of
the coupling strength we have measured the vacuum Rabi splitting of a
single qubit up to the third harmonic resonator mode
(Fig.~\ref{fig:coupling+permutation}(a)) using qubit 2 of sample B (for the
parameters, see Table \ref{tab:parameters}). The simple estimate shows
good agreement with the measured coupling strengths
$g^{(2)}_{0,1,2,3}/2\pi=\{42,\,84,\,125,\,162\}~\rm{MHz}$ (dashed line
in Fig.~\ref{fig:coupling+permutation}(a)). The parameter $\beta=0.20$, obtained from
a linear fit to the analytic model in the large $E_J/E_C$ limit
\cite{Koch2007}, agrees with the designed value within 10\%. A
numerical simulation of the transmon including four energy levels can
explain also the slight deviations from the linear dependence at high
frequencies (solid red line, Fig.~\ref{fig:coupling+permutation}(a)).

In the case of two resonant qubits with identical coupling strength to the
$j$-th harmonic mode, the exchange coupling strength is, according to
Eq.~(\ref{eq:J}), linear in the qubit transition frequency
$\omega_{ge}$ and the mode frequency $(j+1)\omega_0$,
\begin{equation}
\label{eq:gscale} 
J\propto g^{(1)}_jg^{(2)}_j\approx g^2_j \propto
\omega_{ge}\omega_j = (j+1)\omega_{ge}\omega_0.
\end{equation}
When the detuning of the
qubits to the resonator mode is varied, the frequency $(j+1)\omega_0$ of the $j$-th resonator mode is constant and $J$ scales proportional to the
transition frequency $\omega_{ge}$.

\section{Dark state symmetry}
\label{sec:darksymm} 
  \begin{figure}
    \centering
    \includegraphics[width=86mm]{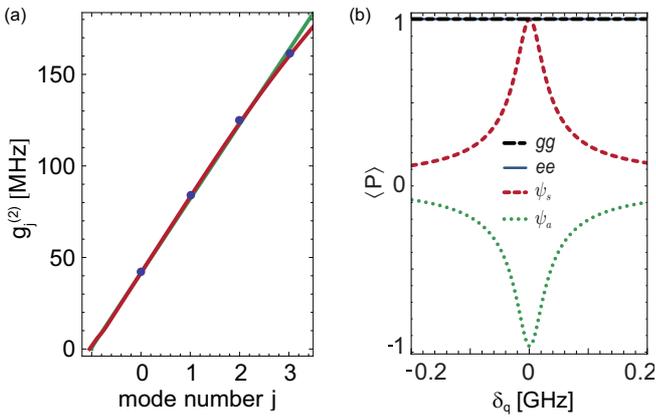} 
    \caption{(a) Coupling strength $g^{(2)}_j$ of qubit 2 to the $j$-th harmonic
mode in sample B. (b) Expectation value of the permutation operator $P$ for the
eigenstates of the Jaynes-Cummings Hamiltonian as a function of
qubit-qubit detuning $\delta_q$.}
    \label{fig:coupling+permutation}
  \end{figure}
The spectroscopic drive $H_d \propto
(\sigma_+^{(1)} - \sigma_+^{(2)}) + h.c.$ \cite{Blais2007}
anti-commutes with the permutation operator 
\begin{equation}
\label{eq:permop}
P \equiv \left(\sigma^{(1)}_+\sigma^{(2)}_- +
   \sigma^{(2)}_+\sigma^{(1)}_-\right) + \left(1 +
\sigma_z^{(1)}\sigma_z^{(2)}\right)/2,
\end{equation}
\begin{equation}
\left[H_d,P\right]_+ \equiv H_d P
+ P H_d = 0.
\end{equation}
 This relation can be fulfilled only if the drive transforms a
symmetric ($\psi_{s}$) to an anti-symmetric ($\psi_{a}$) state, or
vice versa, such that
 \begin{align} \left[H_d,P\right]_+ \psi_{s} &= H_d P \psi_{s} + P
H_d \psi_{s}\\\nonumber & = H_d \psi_{s} + P \psi_{a}\\\nonumber &= \psi_{a} - \psi_{a}
= 0.
 \end{align} 
The symmetry of the coupled qubit states (see Figure \ref{fig:2Qsplitting}(a)) can be characterized by the corresponding expectation value of the permutation operator $\langle P\rangle$.
$\langle P \rangle$ is one for the $\ket{gg}$ and the $\ket{ee}$-state, i.~e. the $\ket{gg}$ and
$\ket{ee}$ states are symmetric for all detunings. For the symmetric and
anti-symmetric states $\psi_s$ and $\psi_a$ formed at zero qubit-qubit detuning, $\langle P \rangle$ is
$1$ or $-1$ indicating that these states
are eigenstates of ${P}$ with well-defined symmetry. 
For non-zero detuning between the qubits, $\delta_q \neq 0$,
the eigenstates of the Hamiltonian (\ref{eq:HJ}) do not have well-defined symmetry and  $\langle P \rangle$ approaches
asymptotically zero for large detunings (Fig.~\ref{fig:coupling+permutation}(b)). 
Hence, no strict
selection rules are imposed off the level crossing and the
transition between ground state and single excited states is allowed.



\end{document}